\documentclass[useAMS,usenatbib]{mn2e}
\usepackage{graphicx}
\title[New O2\,If*/WN6 stars in Westerlund 2]{Two O2\,If*/WN6 stars possibly ejected from the massive young
 Galactic cluster Westerlund 2}
\author[A. Roman-Lopes, R. H. Barba \& N. I. Morrell]{A. Roman-Lopes$^{1}$\thanks{E-mail:
roman@dfuls.cl}, R. H. Barba$^{1,2}$ \& N. I. Morrell$^{3}$\\
$^{1}$Department of Physics - Universidad de La Serena - Benavente, 980 - La Serena - Chile\\
$^{2}$Instituto de Ciencias Astronomicas, de la Tierra, y del Espacio (ICATE-CONICET), Av. Espa\~na 1512 Sur, J54202DSP,\\
San Juan, Argentina\\
$^{3}$Las Campanas Observatory, Carnegie Observatories, Casilla 601, La Serena, Chile }

\begin{document}

\date{}

\pagerange{\pageref{firstpage}--\pageref{lastpage}} \pubyear{2010}

\maketitle

\label{firstpage}

\begin{abstract}
In this paper we report the identification of two new Galactic O2\,If*/WN6 stars (WR20aa and WR20c), 
in the outskirt of the massive young stellar cluster Westerlund 2.
The morphological similarity between the near-infrared spectra of the new stars with that of 
WR20a and WR21a (two of the most massive binaries known to date) 
is remarkable, indicating that probably they are also very massive stars.
New optical spectroscopic observations of WR20aa suggest an intermediate O2 If*/WN6 spectral type.
Based on a mosaic made from the 3.6$\mu$m Spitzer IRAC images of the
region including part of the RCW49 complex, we studied the spatial location
of the new emission line stars, finding that
WR20aa and WR20c are well displaced from the centre of Westerlund 2, 
being placed at $\approx 36$ pc (15.7 arcmin)
and $\approx 58$ pc (25.0 arcmin) respectively, for an assumed heliocentric distance of 8 kpc.
Also very remarkably, a radius vector connecting both
stars would intercept the Westerlund 2 cluster exactly at the place
where its stellar density reaches a maximum.
We consequently postulate a scenario in which WR20aa and WR20c had a
common origin somewhere in the cluster core, being ejected from their
birthplace by dynamical interacion with some other very massive objects, perhaps during some 
earlier stage of the cluster evolution.

\end{abstract}


\section{ Introduction}

\begin{table*}
\begin{center}
\caption{Coordinates and Optical/NIR photometry of the newly-identified O2\,If*/WN6 stars in the outskirt of Westerlund 2.\label{tbl-2}}
\renewcommand{\thefootnote}{\thempfootnote}
\begin{tabular}{ccccccccc}
\\
\hline\hline
Star & RA     &  Dec & $B$  & $V$ & $J$ & $H$ & $Ks$ & Other designation \\
~ &(J2000) & (J2000) &~    & \\
\hline
WR20aa          &10:23:23.49  & -58:00:20.8  &   13.86$\pm{0.02}$  &   12.69$\pm{0.02}$  & 9.28$\pm{0.02}$ & 8.73$\pm{0.02}$ & 8.40$\pm{0.02}$ & SS215 \\
WR20c           &10:25:02.60  & -57:21:47.3  &   20.10$\pm{0.05}$  &   17.51$\pm{0.03}$  & 10.51$\pm{0.02}$ & 9.57$\pm{0.02}$ & 9.04$\pm{0.02}$ & ---  \\
\hline

\end{tabular}
\end{center}
\end{table*}

   \begin{figure*}
   \centering
   \includegraphics[bb=14 14 562 484,width=15 cm,clip]{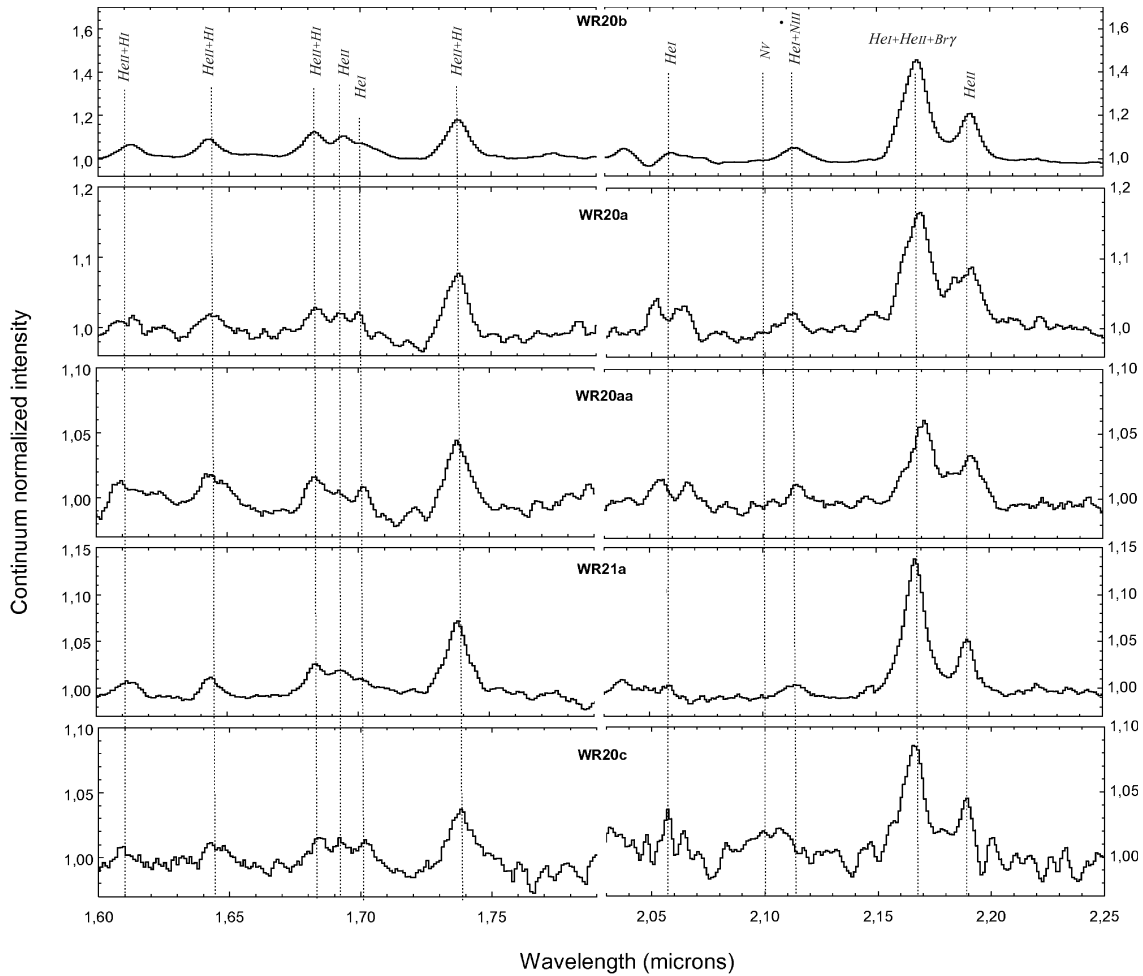}
      \caption{The H- and K-band continuum normalized ESO-NTT-SofI archival spectra of the stars WR20a (O3\,If*/WN6+O3\,If*/WN6), WR20b (WN6ha), 
WR21a (O3\,If*/WN6+early O) together with that of the new stars WR20aa and WR20c. 
The main H, He and N emission lines are identified by labels.}
         \label{FigVibStab}
   \end{figure*} 

Westerlund 2 (Wd2, \citet{b27}, \citet{b26}) is a young massive
stellar cluster placed in the core of the H{\sc ii} region RCW49, which is
located in the fourth Galactic quadrant, projected in the area of the Carina
constellation. 
This cluster hosts one of the most massive binaries known: the star WR20a,
comprised by two O3\,If*/WN6 type stars with absolute masses of 83M$_\odot$ and
82M$_\odot$ \citep{b3,b19}. 
Another massive star of the same spectral-type, WR20b \citep{b29} appears about 4' south-east 
of the cluster core. 
A second massive WR binary, WR21a (that is located 16' east of the Wd2 core) has a
primary component of spectral type O3\,If*/WN6 with an estimated minimum mass of
87M$_\odot$, and an early O type secondary with a minimum mass of
53M$_\odot$ \citep{b1}. However the membership of WR21a to Wd2 is
still uncertain. 
Such WR stars are members of the WNH type, a subset of very
luminous and massive Hydrogen core burning WN objects (see definition in
\citet{b2} and references there-in). 

\begin{table*}
\caption{Journal of the spectroscopic data used in this work.}
\label{catalog}
\begin{tabular}{ccccccccc}
\\
\hline 
\hline
Star & Program ID &  Date     & Telescope & Instrument  &  Grism/Disperser   &   Slit  &  R  & Range ($\mu$m) \\
\hline
WR20a & 075.D-0210 & 2005-06-21  & NTT       &    SofI     &  GR       &
0.6$\arcsec$ x 290$\arcsec$ & 1000 & 1.53-2.52  \\ 
      & 072.D-0082 & 2004-01-29  & NTT       &    EMMI/2.3     &  GRAT\#3
      & 1$\arcsec$ x 120$\arcsec$ & 1600 & 0.39-0.47  \\ 
WR20b & 075.D-0210 & 2005-06-20  & NTT       &    SofI     &  GR       &  0.6$\arcsec$ x 290$\arcsec$ & 1000 & 1.53-2.52  \\
WR21a & 075.D-0210 & 2005-06-20  & NTT       &    SofI     &  GR       &  0.6$\arcsec$ x 290$\arcsec$ & 1000 & 1.53-2.52  \\
WR20aa & 075.D-0210 & 2005-06-21  & NTT       &    SofI     &  GR       &  0.6$\arcsec$ x 290$\arcsec$ & 1000 & 1.53-2.52  \\
      & GOSSS & 2010-06-30  & du Pont      &    Boller \& Chivens  &  1200 lines mm$^{-1}$       & 1$\arcsec$x270$\arcsec$ & 2500 & 0.39-0.55  \\
WR20c & 075.D-0210 & 2005-06-21  & NTT       &    SofI     &  GR       &  0.6$\arcsec$ x 290$\arcsec$ & 1000 & 1.53-2.52  \\
WR25 & GOSSS & 2008-05-20  & du Pont      &    Boller \& Chivens  &  1200 lines mm$^{-1}$     & 1$\arcsec$x270$\arcsec$ & 2500 & 0.39-0.55  \\
HD93129A & GOSSS & 2010-06-30  & du Pont      &    Boller \& Chivens  &  1200 lines mm$^{-1}$     & 1$\arcsec$x270$\arcsec$ & 2500 & 0.39-0.55  \\
\hline  
\hline
\end{tabular}
\end{table*}

With empirical masses exceeding 80 M$_\odot$, WNH stars are probably among the
most massive stars known, being  in the top-end of the mass distribution in
young clusters and OB associations. This assumption is supported by results
obtained from several studies of WNH stars in massive clusters such as R145 in 30
Doradus \citep{b20}, WR25 \citep{b22} and
WR22 \citep{b23} in the Carina Nebula region, and NGC3603-A1 in NGC3603 \citep{b21}. 

More recently, \citet{b14} from the spectroscopic analyses of four WN
stars located within R136 (the core of 30 Doradus Nebula in the Large Magellanic Cloud - LMC) and
three WN stars of HD 97950 (the core of NGC 3603 in the Milky Way), computed 
initial masses in the range $165-320$ M$_\odot$, and $105-170$ M$_\odot$, 
for the stars in R 136 and NGC 3603, respectively. 
They point out that due to their proximity to the Eddington limit, the very
high-mass progenitors would possess an emission-line spectrum, 
at the beginning of their main-sequence evolution, mimicking the
spectral appearance of classical WR stars. 

The spectral-type O3If*/WN6-A was introduced almost
thirty years ago by \citet{b28} to classify the bright emission line 
star Sk-67 22 in LMC, which shows an intermediate spectrum between 
those of HD93129A (O2\,If*) and HD93162 (=WR25, O2\,If*/WN6).
New examples of this intermediate spectral type were mainly
found in LMC, always consisting in bright massive objects (see
for example, \citet{b35}, \citet{b36}, \citet{b14}).

   \begin{figure*}
    \vspace{0pt}
   \centering
   \includegraphics[bb=14 24 580 514,width=12 cm,clip]{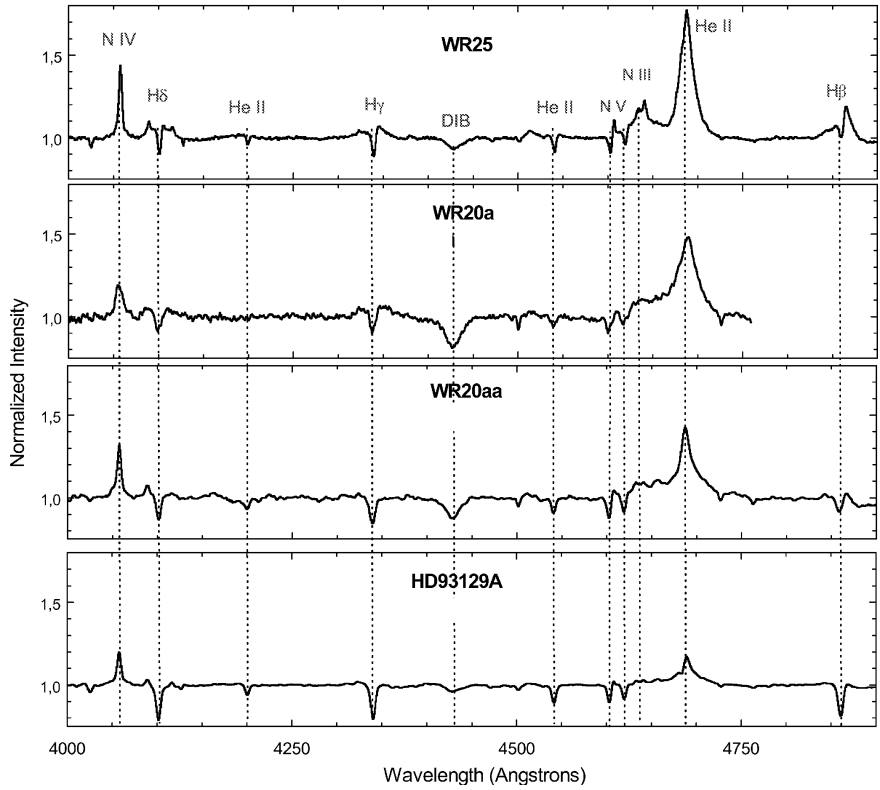}
      \caption{The optical spectrum of WR20aa, together
with that for HD 93129A (O2\,If*), WR 25 (O2.5\,If*/WN6 + O), and WR20a (O3\,If*/WN6 + O3\,If*/WN6). 
A comparison with the spectrum of HD93129A and that for WR20a shows that the optical
spectrum of WR20aa approximates better to the later.}
         \label{FigVibStab}
   \end{figure*} 

\begin{table*}
\caption{Equivalent widths (\AA\ ) of the main emission lines detected in the NTT-SOFI H- and K-band spectra of WR20aa, WR20c 
WR20a, WR20b and WR21a.}
\label{catalog}
\centering
\renewcommand{\footnoterule}{}  
\begin{tabular}{ccccccccccc}
\hline \hline
& Line  & He{\sc ii}+Br12 &He{\sc ii}+Br11 &He{\sc ii}  &He{\sc i}  & He{\sc ii}+Br10 & He{\sc i}+N{\sc iii} & He{\sc i}+He{\sc ii}+Br$\gamma$ &He{\sc ii} \\
\hline
& $\lambda$ ($\micron$)  & 1.645  & 1.685 &  1.693   & 1.702 & 1.738 & 2.115 & 2.167 & 2.190 \\
\hline
& WR20b (WN6ha)  & -7.4 $\pm$1.5 & -11.8 $\pm$2.1 & -5.3 $\pm$1.0 & -7.0 $\pm$0.8 & -18.6 $\pm$2.1 & -12.1 $\pm$1.9  & -74.5 $\pm$10.6  & -33.7 $\pm$4.5 \\
& WR20a (O3\,If*/WN6+O3\,If*/WN6) & -3.0 $\pm$0.7 & -3.3 $\pm$0.7 & -1.7 $\pm$0.4 & -1.3 $\pm$0.3 & -8.0 $\pm$1.4 & -2.0 $\pm$0.5  & -29.9 $\pm$6.3  & -14.5 $\pm$3.8 \\
& WR21a (O3\,If*/WN6+O) & -1.5 $\pm$0.3 & -2.3 $\pm$0.5 & -2.9 $\pm$0.7 & -1.9 $\pm$0.5 & -10.2 $\pm$2.4 & -1.2 $\pm$0.5  & -20.1 $\pm$5.2  & -5.3 $\pm$1.3 \\
& WR20aa (O2If*/WN6)& -3.1 $\pm$1.0 & -3.3 $\pm$1.0 & -0.2 $\pm$0.1 & -1.6 $\pm$0.4 & -6.1 $\pm$1.3 & -1.1 $\pm$0.4  & -9.9 $\pm$2.1  & -4.8 $\pm$1.2 \\
& WR20c (O2If*/WN6) & -1.5 $\pm$0.5 & -1.5 $\pm$0.6 & -1.3 $\pm$0.5 & -1.3 $\pm$0.4 & -5.2 $\pm$1.2 & -2.7 $\pm$0.7  & -12.5 $\pm$2.7  & -4.0 $\pm$1.2 \\
\hline
\end{tabular}
\end{table*} 

In this paper we report the identification of two new Galactic O2\,If*/WN6 stars (WR20aa
and WR20c) in the outskirt of the massive young cluster Wd2.  
In section 2 we describe the used observational data and the data reduction
procedures, in section 3 we present the results and discussion, and in section
4 it is performed a summary of the work.

\begin{table*}
\caption{Equivalent widths (\AA\ ) of the main diagnostic lines detected in the Blue optical spectra of WR20aa, HD93129A, 
WR20a, and WR25 stars.}
\label{catalog}
\centering
\renewcommand{\footnoterule}{}  
\begin{tabular}{cccccc}
\hline \hline
 ID  & N{\sc iv} 4058 &He{\sc ii} 4686  &H$_\beta$  & N{\sc v} 4604 & N{\sc iii} 4634-40-42 \\
\hline
 WR25  & -2.5 $\pm$0.1 & -15.6 $\pm$0.7 & -2.7 $\pm$0.1 & 0.3 $\pm$0.1 & -0.4 $\pm$0.1 \\
 WR20a  & -2.3 $\pm$0.1 & -14.0 $\pm$0.6 & --- & 0.4 $\pm$0.1 & -0.2 $\pm$0.1  \\
 WR20aa  & -2.4 $\pm$0.1 & -9.6 $\pm$0.5 & 0.3 $\pm$0.1 & 0.7 $\pm$0.1 & -0.2 $\pm$0.1 \\
 HD93129A & -1.3 $\pm$0.1 & -3.4 $\pm$0.4 & 1.3 $\pm$0.1 & 0.5 $\pm$0.1 & -0.1 $\pm$0.1 \\
\hline
\end{tabular}
\end{table*}

\section{Observations and data reduction}

\subsection{Optical and Near-Infrared photometry of WR20aa and WR20c}

Coordinates and photometric parameters for the new O2\,If*/WN6 stars are shown in
Table 1. The near-IR photometric data are from the Two-Micron All Sky Survey
(2MASS, \citet{b24}), and were retrieved using the
NASA/IPAC\footnote{http://irsa.ipac.caltech.edu/applications/BabyGator/}
Infrared Science Archive. 
Bessell B- and V-band imaging of WR20aa and WR20c was obtained with the
Swope 1-m telescope at Las Campanas Observatory (Chile) using the SITe 3 CCD
detector, on April 20, 2010. The night was photometric, with a typical seeing of 1.2
arcsec. 
The data were reduced following standard IRAF\footnote{http://iraf.noao.edu/} procedures. We then obtained
aperture photometry for both stars, and
their instrumental magnitudes were transformed to the standard system using a
set of photometric standard stars observed during the related night.

The stars were named WR20aa and WR20c, following the common practice consisting 
in giving the WR Galactic stars
numbers according to their RA, with further additions  
between integers following \citet{b15,b16}. Searching in 
SIMBAD\footnote{http://simbad.u-strasbg.fr/simbad/}, 
we found that WR20aa is cataloged as SS\,215, an H$\alpha$ emission-line object discovered
by \citet{b7}.

\subsection{Spectroscopic observations}

\subsubsection{Near-Infrared spectroscopy}

Near-Infrared (NIR)
ESO\footnote{http://archive.eso.org/eso/eso\_archive\_main.html} archival
spectra obtained with the SofI instrument \citep{b5}, coupled to the 3.5 m New
Technology Telescope (NTT) are part of the dataset used in this work. 
These spectra were taken as part of the ESO program 075.D-0210 (PI Marston), with the
targets being selected accordingly to near- to mid-IR colour criteria for stars possessing strong winds \citep{b4}.  
The log file of the NIR spectroscopic dataset is presented in
Table 2.

\begin{table*}
\begin{center}
\caption{(B-V) color, color excess, visual extinction and absolute visual magnitudes for WR20aa, WR20c, WR20a, and WR20b. 
Data for WR21a are included as a complement.  In the entry labeled
\textit{Early O members}, we present the interval of (B-V) colors, color
excess and visual extinction  for the
12 known members of Wd2. The last
column shows the equivalent single star magnitudes which follow from the consideration
that WR21a and WR20a are binary systems (see text).
 \label{tbl-2}}
\renewcommand{\thefootnote}{\thempfootnote}
\begin{tabular}{cccccc}
\\
\hline\hline
Star & $(B-V)$  & $E(B-V)$ & $A_V$ & $M_V$ & $M_V (corr)$ \\
\hline
WR20aa (O2\,If*/WN6)     &  1.2 &   1.5  & 4.7 & -6.5 & -6.5 \\
WR20c  (O2\,If*/WN6)    &  2.6 &   2.9  & 9.0 & -6.1 & -6.1 \\
WR20a  (O3\,If*/WN6+O3\,If*/WN6)   &  1.6 &   1.9  & 5.9 & -7.0 & -6.3 \\
Early-O members   &  1.2 - 1.6 & 1.5 - 1.9   & 4.7 - 5.9 & --- &  --- \\ 
WR20b  (WN6ha)    &  1.5 &   1.8  & 5.6 & -6.6 & -6.6 \\
WR21a  (O3\,If*/WN6+O)    &  1.4 &   1.7  & 5.3 & -7.2 & -6.7 \\

\hline

\end{tabular}
\end{center}
\end{table*}

The raw frames were reduced following the NIR reduction procedures presented
by \citet{b6}, and briefly described here. 
The two-dimensional frames were sky-subtracted for each pair of images taken
at the two nod positions A and B, followed by division of the resultant image
by a master flat.  
The multiple exposures were combined, followed by one-dimensional extraction
of the spectra. 
Thereafter, wavelength calibration was obtained using the IDENTIFY/DISPCOR
IRAF tasks applied to a set of OH sky line spectra (each with about 35 sky
lines in the range 15500\AA\ -23000\AA\ ). 
The typical error (1-$\sigma$) for this calibration process is estimated as
$\sim$20\AA\, which corresponds to half of the mean FWHM of the OH lines in
the mentioned spectral range. 
Telluric atmospheric corrections were done using H- and K-band spectra of B
type stars obtained before or after the target observation. 
The photospheric absorption lines present in the high signal-to-noise telluric
spectra, were subtracted from a careful fitting (through the use of Voigt and
Lorentz profiles) to the hydrogen and helium absorption lines (He
absorption lines are sometimes seen at 1.70$\mu$m and 2.11$\mu$m in the
earliest B-type stars), and respective adjacent continuum.   

\subsubsection{Optical Spectroscopy}

Optical spectroscopic CCD observations were performed with the Boller \& Chivens
spectrograph attached to the 2.5\,m du Pont telescope at Las Campanas
Observatory (Chile) in June 2010. 
The spectrum was obtained in the framework of the Galactic O-stars Spectroscopic
Survey (GOSSS, \citet{b37}), and therefore using its instrumental setup. A
spectrum of HD\,93129A was also obtained in June 2010, and a spectrum of
HD\,93162 (WR25) in May 2008, during the first GOSSS Southern campaign. 
Additionally, two optical spectra of WR20a were retrieved from the ESO
Archive under the program 072.D-0082 (PI Rauw). 
These spectra were obtained with EMMI instrument attached to the ESO NTT at
La Silla (Chile) in February 2004, and they belong to the same dataset analyzed and
published by \citet{b19}. Data were reduced and normalized using
ONEDSPEC IRAF routines.  
A journal of the optical spectroscopic observations is presented in Table 2. 

\section{Results and Discussion}

\subsection{NIR spectra of WR20aa and WR20c}

In Figure 1 we present the telluric corrected (continuum normalized) H- and
K-band spectra of the WR20aa and WR20c stars, together with those of WR20a, 
WR20b and WR21a. 
The strongest features are the blends of H{\sc i} and He{\sc ii} emission lines
at 1.736$\mu$m and 2.167$\mu$m, as well as the He{\sc i}+N{\sc iii} and 
He{\sc ii} emission lines at 2.115$\mu$m and 2.189$\mu$m, respectively.   
A list with the main emission lines and corresponding equivalent line-widths is
presented in Table 3.

The similarity between the NIR spectra of the known O2\,If*/WN6 stars with that of
WR20aa and WR20c (Figure 1) is remarkable, indicating that the later 
may also belong to the O2\,If*/WN6 spectral type. 
This assumption is reinforced by the close morphological match of the
WR20aa H- and K-band spectra with that for WR20a, one of the most massive
binaries known to date (O3\,If*/WN6+O3\,If*/WN6) for which masses of 83M$_\odot$ and 82M$_\odot$ were derived \citep{b3,b32},
and by the match of the WR20c H- and K-band spectra
with that of WR21a, another extremely massive binary system (O3\,If*/WN6 + 
early O), for which \citet{b1} estimated minimum masses
of 87M$_\odot$ and 53M$_\odot$, respectively.

\citet{b39} showed that some O4\,If stars (like HD\,16691 and 
HD\,190429) can be erroneously classified as WN stars if observed only in the K-band. 
In fact, their K-band emission features 
are albeit morphologically identical to those seen in the spectra presented in 
Figure 1.
However, as can be seen in \citet{b40}, at least in the case of HD\,190429 this is 
not true for the H-band.
Indeed, the H-band spectrum of HD\,190429 is completely different of those
shown in Figure 1, being virtually featureless in the mentioned spectral range.
Also \citet{b38} present the H-band spectrum of Cyg OB2 \#7 
(O3\,If*) which shows several H and He lines in \textit{absorption}.
On the other hand, as can be noticed from Figure 1 the H-band spectra
of WR20aa and WR20c are \textit{not} featureless, with all H and He relevant
lines clearly seen in \textit{emission}.
In this sense, H- and K-band spectra of heavily reddened massive star candidates may became useful tools to discriminate between 
the OIf* and OIf*/WN spectral types.

\subsection{The optical spectrum confirms: WR20aa is an O2\,If*/WN6 star} 


The spectral morphology derived from NIR spectrograms of one of the new sources (WR20aa), is confirmed from new optical 
spectroscopic observations. In Figure 2, we present the optical spectrum of WR20aa, together with GOSSS spectra of HD 93129A 
(O2 If*) and WR25 (O2.5\,If*/WN6 + O) obtained with the same instrumentation and setup, along with the archival ESO/EMMI spectrum of 
WR20a (O3\,If*/WN6 + O3\,If*/WN6) \citep{b46}. Table 4 lists the equivalent widths (EW) for the main diagnostic lines used 
for spectral classification of WR20aa, HD 93129A, WR20a, and WR25.

The blue spectrum of WR20aa is dominated by He{\sc ii} 4686\AA\ and
N{\sc iv} 4058\AA\ emission lines, together with less pronounced emission
lines of N{\sc iii} 4634-40-42\AA\ and Si{\sc iv} 4089\AA\ .  
Numerous absorption lines of H{\sc i} and He{\sc ii}, and also some diffuse interstellar bands 
(DIBs) at $\lambda\lambda$ 4428\AA\, 4726\AA\, 4765\AA\, and 4865\AA\ are present. A special mention deserve the strong 
N{\sc v} 4604-20\AA\ absorptions, which are characteristic of the earliest O-type (O2-3) and WNha stars. These absorption lines show 
weak P-Cyg profile structure. The  strength of the N{\sc iv} 4058\AA\ emission line relative to the N{\sc iii} 4634-40-42\AA\ emission lines  
indicates that the spectral type of WR20aa is at least as early as O2, according to the criteria described in \citet{b43}. 
In addition, the presence of a weak P-Cyg profile in the H-beta line of WR20aa, compared to the same line in HD\,93129A, 
which shows no emission, and the 
P-Cyg profile observed in  WR25, is indicative of an intermediate spectral type between  OIf* and WNha. 
In this way, \citet{b42} suggest that genuine WN6 stars are those with EW of He\,{\sc ii} 4686\AA\ lower than -12\AA\ . 
The measurement of He\,{\sc ii} 4686\AA\ line in WR20aa (-9.6\AA\, see Table 4) also indicates an intermediate spectral type. 
On the other hand, no detection of He\,{\sc i} absorptions is also indicative of a spectral type earlier than the O3. 
This picture is seen in the direct comparison with the spectra of HD\,93129A and WR20a, which shows that the blue
spectrum of WR20aa resembles better to the later (as it was already noticed from the NIR regime). It is also noticeable 
from Table 4 that the emission lines of WR20aa are closer in strength to those from WR20a, being the lines from the later 
slightly broader (this difference is perhaps related to the binary nature of WR20a). Consequently, we classify 
the spectrum of WR20aa as O2If*/WN6 \citep{b46}. 

A spectral classification for WR20c using optical MK criteria is not possible in this moment due to the lack of a 
good signal-to-noise optical spectrum of this star. From inspection of our NIR H- and K-band spectra, we find that WR20c 
is spectroscopically very similar to WR20aa, consequently we propose that WR20c should also be classified as O2\,If*/WN6.

   \begin{figure*}
    \vspace{2pt}
   \centering
   \includegraphics[bb=14 14 558 641,width=15 cm,clip]{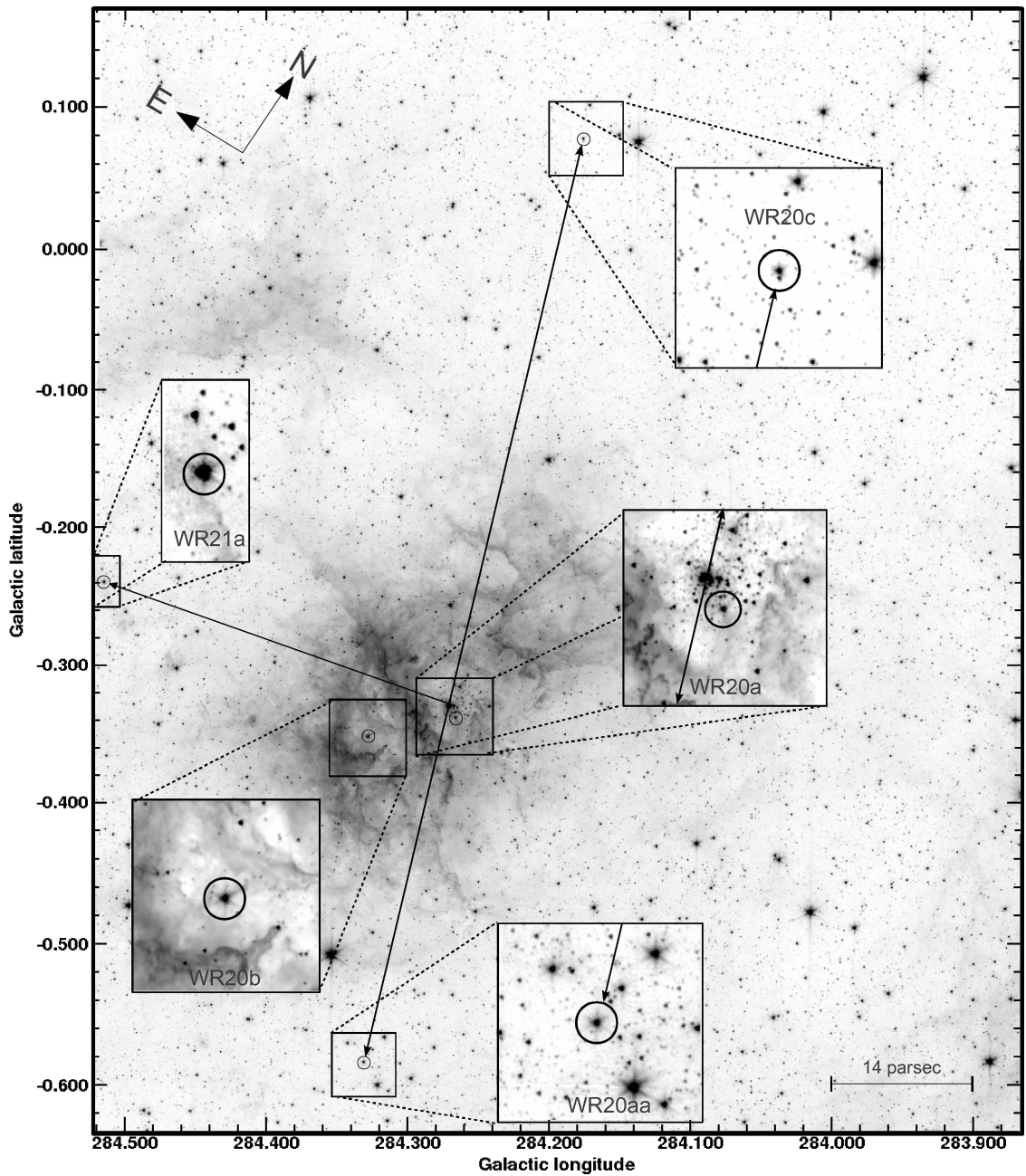}
      \caption{A mosaic made from the 3.6$\mu$m Spitzer IRAC images
(taken from the GLIMPSE survey archival data) of the region towards the RCW49
complex (6 arcmin $\approx$ 14 parsecs at an heliocentric distance of 8 kpc). Here we identify the stars WR20a, WR20b and WR21a 
together with the two newly found O2\,If*/WN6 objects, WR20aa and WR20c.
We notice that a line connecting WR20aa and WR20c
intersects the cluster centre just where the surface stellar density hits
a maximum..}
         \label{FigVibStab}
   \end{figure*} 

\subsection{The new O2\,If*/WN6 stars and the Westerlund 2 cluster: are they related?}

In Figure 3 it is shown a mosaic made from the 3.6$\mu$m Spitzer IRAC images \citep{b33} of the region towards the RCW49
complex. 
There we identify WR20a, WR20b and WR21a together with WR20aa and WR20c. 
It is interesting to notice that the three already known WR stars are well
displaced from the core of the Wd2 cluster, which harbors at least a dozen of early O-type 
stars \citep{b10}.
The closest of them (WR20a), is located at about 0.6 arcmin from
the cluster centre, with the others being placed at core distances of about 3.7 arcmin and 
15.6 arcmin, respectively.
The situation for WR20aa and WR20c is similar to that of WR21a, with 
the stars localized at angular distances (from the cluster centre) of 15.7 arcmin
and 25 arcmin, well beyond the Wd2 cluster core.
The presence of such massive stars well isolated in the outskirt of the RCW49 complex, naturally
leave us to ask: how did they arrive there?  

\subsubsection{Interstellar extinction and absolute visual magnitudes }

The assumption of a possible physical connexion between WR20aa and WR20c with the Wd2 cluster population,
in principle can be tested by comparing the (B-V) colors of these new stars 
with those of
the presently accepted cluster members (WR20a and the 12 known early O-type members),
which we do in Table 5. 
For the sake of completeness, we also 
included here the colors of WR20b and WR21a, the other two known WR stars in the vicinity of Wd2.
The (B-V) colors of WR20aa and WR20c were computed from the B- and 
V-band photometry shown in Table 1, while the values for the
other stars were taken from \citet{b10} (WR20a and the early-O stars), 
\citet{b30} (WR20b), and \citet{b1} (WR21a). In the case of the
early O-stars, we indicate the interval of (B-V) colors shown 
in Table 1 of \citet{b10}. 
The corresponding visual extinctions were computed considering intrinsic 
colors (B-V)$_0$= -0.3 magnitudes, and the
canonical value for the ratio of total to selective extinction R$_V$=3.1. Finally, 
absolute visual magnitudes were estimated
assuming a distance of about 8 Kpc \citep{b31,b10}.

In case of WR20aa we notice that its derived (B-V) color and visual extinction are compatible with those 
for the other massive stars in Wd2, while 
a very red color index (B-V)=2.6 was obtained for WR20c, which corresponds to about 
twice of the visual extinction inferred for the early-O stars of Wd2.
The higher interstellar extinction derived for WR20c is probably due to the presence of a foreground molecular cloud as can 
be inferred from an inspection of the 2MASS J-band source density map\footnote{http://aladin.u-strasbg.fr/java/nph-aladin.pl} 
of the region towards Wd2. Also, we remember that in the case of WR20c, the star is almost on the Galactic Plane 
which increases the probability that a dusty screen is on the line-of-sight of this star. 
In fact, it is not unusual that star forming regions present highly variable visual 
extinctions at scales of a few arcminutes. As examples, we can mention M17 \citep{b44} and the Cygnus OB2 association \citep{b45}
in which the massive stars present a 
wide range (A$_V$\,=\,5-12 and A$_V$\,=\,5-20 magnitudes respectively) of individual visual absorptions.

On the other hand, the absolute visual magnitudes derived for WR20aa 
and WR20c are similar to that for WR20b, which 
is the only one in
the area of RCW49, not known to be a binary.
Indeed, if we 
correct the absolute visual magnitudes 
of WR20a (O3\,If*/WN6+O3\,If*/WN6) and WR21a (O3\,If*/WN6+
early-O) in order to take into account the binary companion contribution (by 0.7 and 
0.5 magnitudes, respectively), we can see that WR20aa is 
probably as luminous as the other OIf*/WN or WN6ha 
stars in Wd2, with WR20c being a bit less luminous than them. 
This is not surprising if one considers that stars of same specific spectral types 
may span a range in absolute magnitudes (see Table 3 of \citet{b46}).

\subsubsection{Two runaway stars?}

It is an observational fact that very massive stars preferentially are found in
the core of their parental clusters, normally forming binary or multiple
systems.
However, there are a certain number of very young and massive stars that are
found well isolated in the field.
Two canonical scenarios try to explain the origin of these stars. 
The first known as the \textit{binary-supernova scenario}, considers the
disruption of a short-period binary system from the asymmetric supernova
explosion of one of the binary components,
while the second, named \textit{dynamical ejection scenario}, involves dynamical
three- or multiple-body encounters in dense stellar systems.

The time scale for the binary-supernova scenario involves values larger than the
expected age (1-2 Myrs) for WN6ha stars \citep{b14} although lower mass counterparts 
could reasonably be rather older (approx. 2.5 Myr).
On the other hand, the multiple ejection of massive stars from dense massive 
clusters is
not a new idea (see discussion in \citet{b17}), and recent discoveries of very 
massive runaway stars (e.g. 30 Dor \#16) highlighted the importance of this
scenario in the evolution of massive clusters \citep{b13,b18}.
For more details on the two models see for example the discussion in \citet{b11} and
references there-in. 

We searched for some observational constrain that could give
support to the scenario in which WR20aa and WR20c were
ejected from the Wd2 cluster centre following the dynamical ejection model. We found the surprising result that
a line connecting WR20aa and WR20c (see Figure
3) intercepts the cluster core \textit{exactly} where the star surface density hits a maximum.
Also interesting, a vector conecting WR21a and the Wd2 core (see Figure 3) forms almost a right
angle with the line connecting WR20aa and WR20c.
This apparent configuration strongly resembles the dynamical scenario proposed 
by \citet{b12}
for the origin of AE Aur, $\mu$ Col and the very eccentric binary $\iota$ Ori, 
as runaway stars.
From N-body simulations the authors conclude that these stars were ejected from
their parental cluster as result of binary-binary interactions occurred
in the Trapezium cluster. In this sense, it is remarkable to point out that 
the massive binary system WR21a is known to present a
high eccentricity ($e\sim0.64$, \cite{b1}).

The fine geometrical alignment between WR20aa, 
WR20c and the Wd2 core, suggests that these stars possibly had a 
common origin somewhere in the
central part of the cluster, being ejected from their birthplace (in a 
timescale not greater than their own age of no
more than 2\,Myrs \citep{b14}, 
with minimum recession velocities (projected into the sky plane) of 
18\,km\,s$^{-1}$ and 28\,km\,s$^{-1}$, respectively.
It is also interesting that \citet{b11} performed numerical simulations of a dynamical 
encounter between single massive stars and the
very massive binary WR20a, founding that in a fly-by encounter the average 
recession velocity attained by a 70-80\,M$_\odot$ star could
be quite moderate (less than 30\,km\,s$^{-1}$), being not formally classified as runaway.  

Taking into account that the O2\,If*/WN6 type stars are very rare and the perfect
alignment of the WR20aa-cluster core-WR20c system, we postulate that both
stars probably were formed somewhere in the core of the Wd2 cluster, being
ejected from their birthplace by dynamical interaction with some other very massive star, in some previous
stage of the cluster evolution. 
Of course, further observational (radial velocities, proper motions),
and theoretical dynamical studies are still necessary to properly confirm our assumption.
However, it is our opinion that the impressive alignment seen between such rare stars and the Wd2 cluster core, is in fact a strong 
clue favoring our hypothesis.
In this sense, we speculate that WR20aa and WR20c \textit{may be} two of the most interesting 
Galactic runways known to date.

\section{Summary}

In this work we report the detection of two new Galactic O2\,If*/WN6 stars (WR20aa
and WR20c) in the outskirt of the massive young cluster Wd2.

The similarity between the NIR spectra of WR20aa and WR20c with those of WR20a and WR21a (two of the most massive binaries known to date), 
is remarkable, suggesting that they could be members of the O2\,If*/WN6 group. Indeed, the optical spectral morphology indicates that WR20aa presents 
an intermediate O2 If*/WN6 spectral type, based in the intensities of the emission lines of He{\sc ii} 4686\AA\ , N{\sc iv} 4058\AA\ , N{\sc iii}
4634-40-42\AA\ , the N\,{\sc v} absorptions at 4604-20\AA\ , and the P-Cyg profile observed in H$\beta$. As an optical spectrum of
WR20c is not yet available, we propose for it a spectral type similar to
that of WR20a, based on the similarity of their NIR spectra.

From the analysis of the spatial distribution of the new O2\,If*/WN6 stars through the
mosaic made from the 3.6$\mu$m Spitzer IRAC images of the region towards the RCW49 
complex, we found the very interesting result that WR20aa and WR20c are
placed at angular distances of 15.7 arcmin ($\approx 36$ pc) and 25
arcmin ($\approx 58$ pc) from the Wd2 core, respectively. The stars are well 
isolated in the 
outskirt of the RCW49 complex with the radius vector connecting them intercepting 
the cluster core exactly where the star density is maximum. 

Considering the dynamical ejection model, we propose a scenario in
which WR20aa and WR20c had a common origin somewhere in the central
part of Wd2, being ejected from their birthplace in a timescale not
greater than their own age, which is probably no more than 1-2 Myrs.
Taking into account the \textit{rarity} of such massive stars, and considering
the perfect geometrical alignment observed between WR20aa, the cluster core, and WR20c,
we believe that WR20aa and WR20c may represent one of the most interesting
cases supporting the ejection of massive stars, produced by dynamical interaction with other massive companions.

\section*{Acknowledgments}

We would like to thank the anonymous referee by the careful reading of the manuscript. Her/his
comments were valuable to improve the clarity and presentation of the paper.
We also would like to thank Dr. Nolan Walborn for stimulating discussions concerning the
classification of intermediate OIf/WN6 stars. 
This work was partially supported by the ALMA-CONICYT Fund, under the project
number 31060004, \textit{A New Astronomer for the Astrophysics Group - Universidad de
La Serena}, by the Department of Physics of the Universidad de La Serena and by the Research Direction (DIULS) of the Universidad de La Serena. 
This research has made use of the NASA/ IPAC Infrared Science Archive, which
is operated by the Jet Propulsion Laboratory, California Institute of
Technology, under contract with the National Aeronautics and Space
Administration. 
This publication makes use of data products from the Two Micron All Sky
Survey, which is a joint project of the University of Massachusetts and the
Infrared Processing and Analysis Center/California Institute of Technology,
funded by the National Aeronautics and Space Administration and the National
Science Foundation. 
Also, this research has made use of the SIMBAD database, operated at CDS,
Strasbourg, France. 
Partially based on observations made with ESO Telescopes at the La Silla
Observatory under programs IDs $<$072.D-0082(A)$>$, $<$075.D-0210(A)$>$. ARL acknowledge financial support from DIULS through Project CD11103.
RHB acknowledge financial support from DIULS through Regular Project PR10101.

\end{document}